\documentclass[twocolumn]{aastex631}

\newcommand{\rev}[1]{#1}

\received{August 28, 2024}

\submitjournal{AJ}

\shorttitle{Empirical NIRISS Backgrounds}
\shortauthors{Hviding et al.}

\graphicspath{{figures}}

\begin{document}

\title{Improved Empirical Backgrounds for JWST NIRISS Image/WFSS Data Reduction}

\author[0000-0002-4684-9005]{Raphael E. Hviding}
\affiliation{Max-Planck-Institut für Astronomie, Königstuhl 17, D-69117 Heidelberg, Germany}

\author[0000-0003-1665-2073]{Ivelina G. Momcheva}
\affiliation{Max-Planck-Institut für Astronomie, Königstuhl 17, D-69117 Heidelberg, Germany}

\author[0000-0003-1249-6392]{Leonardo Clarke}
\affiliation{Department of Physics and Astronomy, University of California Los Angeles, 430 Portola Plaza, Los Angeles, CA 90095, USA}

\begin{abstract}

The Near Infrared Imager and Slitless Spectrograph (NIRISS) on the James Webb Space Telescope (JWST) is a versatile instrument for collecting imaging and wide-field slitless spectroscopy (WFSS) data for surveys of galaxy clusters, emission-line galaxies, stellar populations, and more. 
Dispersed zodiacal light imprints distinct structures on space-based near-infrared imaging and WFSS observations, necessitating careful subtraction during JWST NIRISS data reduction. 
As of 2024-09-24 NIRISS WFSS calibration backgrounds introduce significant spatially-dependent artifacts, up to 5\% of the overall background level, which can severely affect data quality and following astronomical analysis.
Notably, there are no existing backgrounds for NIRISS imaging data which also show systematic artifacts, such as the `light saber' effect.
In this work, we present improved empirical JWST NIRISS imaging and WFSS backgrounds derived from all available public data in the F115W, F150W, and F200W filters.
We demonstrate that our empirical backgrounds provide a more accurate representation of the background structure in NIRISS imaging and WFSS data than existing reference files, mitigating the impact of spatially-dependent artifacts. 
Our empirical backgrounds are publicly available and can be used to improve the quality of JWST NIRISS imaging and WFSS data reduction.

\end{abstract}


\section{Introduction} \label{sec:intro}

The James Webb Space Telescope \citep[JWST;][]{gardnerJamesWebbSpace2006,gardnerJamesWebbSpace2023} represents a monumental leap forward in our ability to explore the universe. 
Equipped with a suite of advanced instrumention, including the Near Infrared Imager and Slitless Spectrograph \citep[NIRISS;][]{doyonJWSTFineGuidance2012,doyonInfraredImagerSlitless2023,willottInfraredImagerSlitless2022}, JWST is capable of capturing images and spectra of astronomical objects in the near- \rev{through} mid-infrared at unprecedented sensitivity and resolution. 
JWST is already making groundbreaking discoveries in the study of early universe and of the formation and evolution of galaxies. 

NIRISS, in particular, is being used to perform extragalactic surveys with imaging and wide-field slitless spectroscopy (WFSS) data \citep{willottInfraredImagerSlitless2022}. 
As of \rev{2024-09-24, 85}\% of the total \rev{science} exposure time on NIRISS has been dedicated to pure-parallel extragalactic surveys, where, during an observation of a primary target, NIRISS is used to observe a secondary field simultaneously.
WFSS enables the simultaneous acquisition of spectra over a wide field, facilitating comprehensive studies of galaxy formation, evolution, and the large-scale structure of the universe. 
The success of programs such as 3D-HST \citep{brammer3DHSTWidefieldGrism2012} has underscored the efficacy of these techniques in yielding critical insights into galaxy evolution.
Upcoming space-based observatories are primed to further advance this domain, with the Euclid \citep{laureijsEuclidDefinitionStudy2011} and Roman Space Telescope \citep{spergelWideFieldInfrarRedSurvey2015} poised to leverage slitless spectroscopy over even larger areas of the sky.

Proper calibration of NIRISS data is therefore crucial for imaging and WFSS surveys to maximize the scientific return of JWST and future observatories.
Accurate calibration is particularly challenging due to factors such as background light, which can introduce spatially varying structures in NIRISS imaging and WFSS data.
These structures require detailed subtraction during data reduction for the subsequent astronomical analysis, such as source detection, photometry, and redshift fitting.
In this study, we address the pressing need for improved empirical calibration backgrounds for NIRISS imaging and WFSS data.

In Section \ref{sec:data} we provide an overview of NIRISS imaging/WFSS data, the shortcomings of calibration backgrounds available as of 2024-09-24, and selection criteria used herein.
Section \ref{sec:method} describes our methodology for developing the new calibration backgrounds, including source mask creation, median combining, and denoising, along with our background subtraction prescription.
In Section \ref{sec:results} we present our results, demonstrating the effectiveness of our approach in reducing artifacts and improving data quality.
Finally, in Section \ref{sec:summary} we summarize our findings and discuss the implications of our work for future JWST NIRISS imaging and WFSS data reduction.

\begin{figure*}[ht!]
    \centering
    \includegraphics[width=\textwidth]{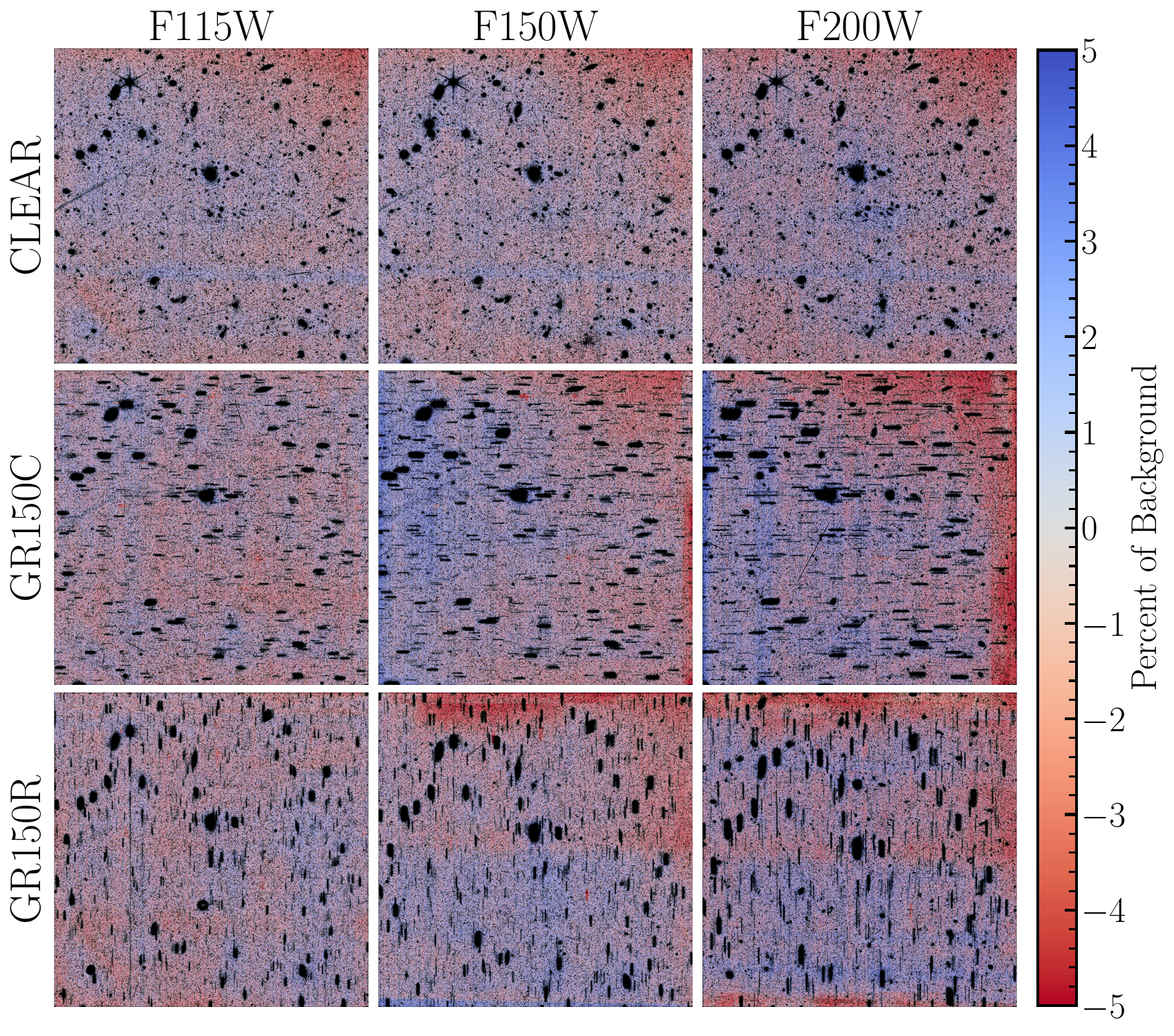}
    \caption{
        Example \rev{NIRISS rate files that have been background-subtracted and flat-fielded} from a specific field across F115W (right column), F150W (middle column), and F200W (left column) filters and CLEAR (top row), GR150C (middle row), and GR150R (bottom row) grisms.
        Background subtraction is performed using the CRDS references for WFSS data \rev{and} a scalar value for imaging data.
        Pixels with an active data quality flag are masked in black along with any detected sources to highlight the residual background structure.
        The resultant image is colored by the deviation from zero as a percent of the total background level.
        Consistent 2D structures are visible in the subtracted images at the 1--5\% level.
        \rev{The light-saber effect can be seen for the direct imaging as a thin horizontal enhancement spanning the length of the detector two thirds from the top.}
        \rev{The largest variation for the WFSS exposures occurs at the detector edges parallel to the dispersion axis due to the diminished contribution of the zodiacal light from outside the field of view.}
        \label{fig:example}
    }
\end{figure*}

\section{NIRISS Data} \label{sec:data}

JWST NIRISS is capable of imaging, WFSS, single-object spectroscopy, and aperture masking interferometry. 
NIRISS provides a field of view of 2.2$'\times$2.2$'$ with a pixel scale of 0.066$''$\,px$^{-1}$ and, in WFSS mode, a spectral resolution of $\mathcal{R}\sim$ 160 over a wavelength range of 0.8--2.2\,$\mu$m.
In this work we focus on imaging and WFSS, the most widely used NIRISS modes, which are particularly relevant to extragalactic observations and pure-parallel surveys.
In this section, we discuss the shortcomings of calibration backgrounds as of 2024-09-24 for NIRISS imaging and WFSS data and describe the selection criteria for the data used in this study.

\subsection{Background Artifacts} \label{subsec:backgrounds}

While many factors contribute to the background in JWST observations, e.g. Milky Way ISM, stray light, etc., the zodiacal light is the dominant source of background light in the near-infrared for the observatory \citep{rigbyHowDarkSky2023}.
The contribution of the zodiacal light depends primarily on the ecliptic latitude of the field and on the wavelength of the observation, with high ecliptic latitudes and redder filters (out to $\sim$3\,$\mu$m) exhibiting lower background levels.
This light imprints distinct structures on NIRISS data, especially in WFSS observations when the zodiacal light is dispersed by the grism.
To this end, commissioning program STL-02, in conjunction with processing scripts available at \cite{willottHttpsGithubCom2024}, produced WFSS background reference files that are available from the JWST Calibration Reference Data System (CRDS) \rev{as of CRDS Context 0918 on 2022-07-08. 
Throughout this work we exclusively compare to this set of calibrations and refer to them as the CRDS reference files.}

In addition, scattered light from the NIRISS susceptibility region introduces a distinct structure in NIRISS Imaging and WFSS data, known as the `light saber' effect, which presents as a thin stripe across the full length of the detector \rev{\citep[\S 5.7.1]{doyonInfraredImagerSlitless2023}}\footnote{See the \href{https://jwst-docs.stsci.edu/known-issues-with-jwst-data/niriss-known-issues\#NIRISSKnownIssues-light-saber}{JWST User Documentation} for more information.}.
If the susceptibility region is dominated by zodiacal light, the light saber signal is $\sim$1\% of the overall background level, though in the presence of a bright star it can be significantly higher.
While existing WFSS backgrounds calibrations can help to mitigate the light saber effect in dispersed data, there no \rev{equivalent calibration or pipeline step for either in-field or stray-light background subtraction for NIRISS imaging data}.

However, CRDS reference files for NIRISS WFSS data introduce significant spatially-dependent artifacts.
In Figure \ref{fig:example} we show example \rev{NIRISS rate files that have been background-subtracted and flat-fielded} across all filter/grism combinations explored in this work.
Examples are drawn from a specific visit for which we have data in all filters and grisms.
The background subtraction is performed using the CRDS references for WFSS data and a scalar value for imaging data as described in Section \ref{subsec:crds}.
The resultant images highlight residual background structures after subtraction which are present across all filter/grism combinations at the 1--5\% level of the total background.
While we only demonstrate the residual for a single field here, these structures are consistent across all NIRISS imaging and WFSS data.
\rev{In general, the largest variation for the WFSS exposures is at the edges of the detector that are parallel to the dispersion axis due to diminished contribution of the zodiacal light from outside the field of view.}
We therefore seek to develop improved empirical backgrounds in this work.

\subsection{Data Selection} \label{subsec:selection}

To build robust empirical backgrounds for NIRISS imaging and WFSS data, we begin by querying all available public JWST NIRISS imaging and WFSS data as of \rev{2024-09-24}. 
This analysis focuses exclusively on the F115W, F150W, and F200W filters, which are the most commonly used NIRISS filters for imaging and WFSS observations. 
In fact, all others filters together \rev{comprise} $<4$\% of the total NIRISS imaging and WFSS exposure time.
Consequently, our data is drawn from the large pure-parallel Cycle 1 and 2 NIRISS programs.
We further restrict our analysis to NIRISS data obtained using the NIS readout pattern in full array mode, though our methodology can be applied to other filter/grism combinations and readout patterns as well.

Additionally, we exclude NIRISS data affected by nearby bright stars.
We ensure this by removing all exposures with a Gaia DR3 \citep{gaiacollaborationGaiaDataRelease2023} G-band magnitude star of $G<9.5$ within 2$'$ of the center of the image.
For all remaining exposures, we download level 2 calibrated, i.e.\ rate file, exposures from MAST. 
Table \ref{tab:data} provides a summary of the number of images per filter/grism combination.
Our dataset comprises 1894 NIRISS exposures across nine grism/filter combinations that are uncontaminated by bright stars and with a consistent readout pattern.
The data described here may be obtained from the MAST archive at\dataset[doi:10.17909/j3y0-5613]{https://dx.doi.org/10.17909/j3y0-5613}.

\begin{deluxetable}{l|c|c|c}
    \label{tab:data}
    \tablecaption{NIRISS Data}
    \tablehead{Grism & \colhead{F115W} & \colhead{F150W} & \colhead{F200W}}
    \startdata
    Clear & 65 & 110 & 409 \\ 
    GR150C & 138 & 285 & 400 \\ 
    GR150R & 255 & 238 & 196 
    \enddata
    \tablecomments{Num.\ images per filter/grism.}
\end{deluxetable}

\begin{figure*}[ht!]
    \centering
    \includegraphics[width=\textwidth]{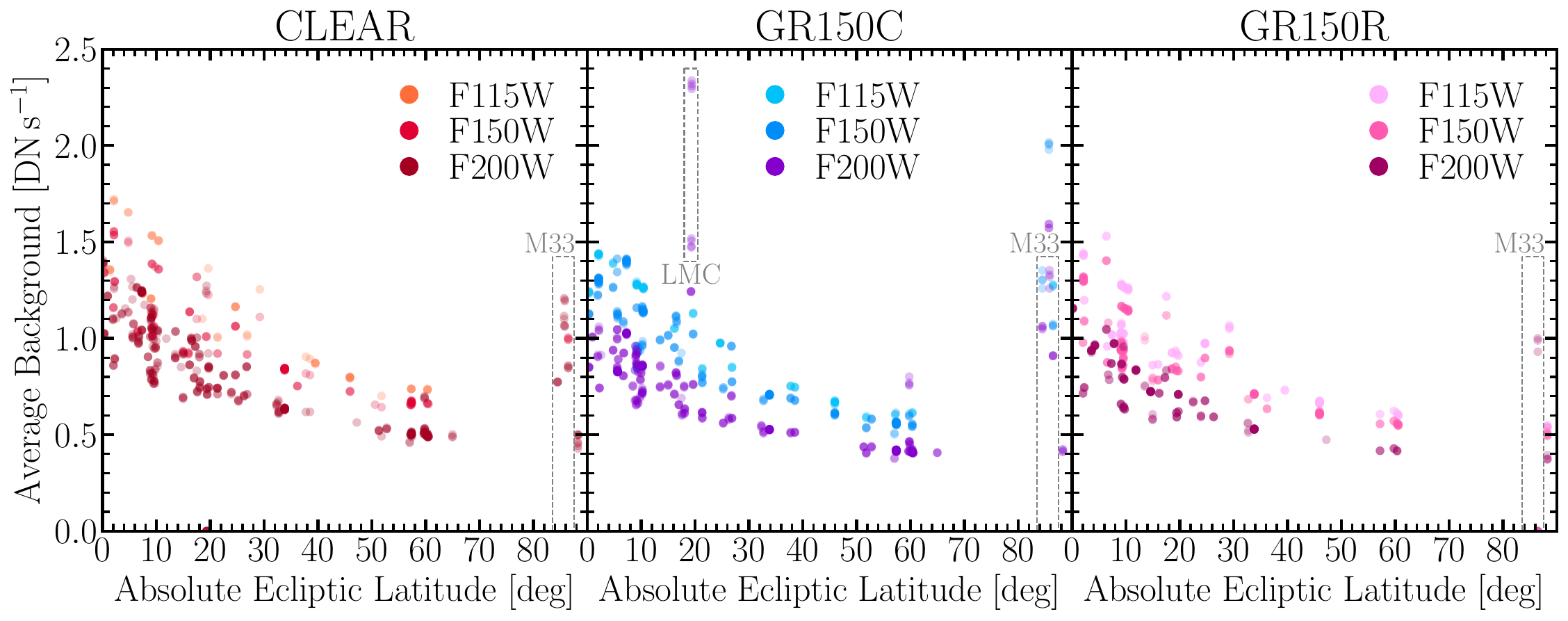}
    \caption{
        Average background level versus absolute ecliptic latitude for CLEAR (left), GR150C (middle), and GR150R (right) grism data in the F115W (light), F150W (medium), and F200W (dark) filters.
        Background levels trend with ecliptic latitude, consistent with the expected zodiacal light contribution.
        In addition, redder filters exhibit a lower background level, consistent with the expected zodiacal light spectrum.
        However, there are several outliers in the data, which are all attributable to observations in dense fields, such as the Large Magellanic Cloud (LMC) and the Triangulum Galaxy (M33), where the background level is dominated by the target itself.
        \label{fig:zodi}
    }
\end{figure*}

\section{Methodology} \label{sec:method}

Before generating improved empirical backgrounds for NIRISS imaging and WFSS data, we first apply standard calibration steps to the rate files.
While background subtraction is done before flat-fielding in the Stage 2 Spectroscopic Processing for WFSS data, we first apply the flat-field to all NIRISS imaging and WFSS data.
This is done to ensure that the flat-fielding process does not introduce any additional artifacts that could affect the empirical background creation.
It should be noted that the backgrounds produced in this work are therefore flat-fielded, as opposed to the WFSS backgrounds used in Stage 2 processing.
We therefore additionally provide a version of the empirical background where the flat-field has been removed to allow for the use of these backgrounds in Stage 2 processing. 
\rev{Furthermore, the flat-field-removed empirical backgrounds provide a convenient way to produce flat-fielded backgrounds should the NIRISS flat-field calibration be updated in the future.}

\subsection{Mask, Combine, \& Smooth} \label{subsec:mask_combine_smooth}

To generate the empirical backgrounds, we start by creating masks for each image in our dataset to identify and exclude sources or artifacts from the background estimation process. 
Initially, we use \texttt{SEP} to estimate a 2D background model for each image. 
Following this, \texttt{photutils} is employed to identify sources with a pixel detection threshold of 2$\sigma$ above the background and a minimum source size of 10 pixels. 
This strict detection threshold ensures that only real sources are masked and not the residual background structure.
In addition, we mask pixels with an active data quality flag, \rev{resulting in a typical masked pixel fraction of $\sim$10\% for each input image.}

The masked images are normalized to the median value of their unmasked pixels and median combined using an iterative sigma-clipping algorithm to generate an estimate of the empirical background for each pixel. 
Pixels with no final value, i.e., those that are masked across all inputs, are subsequently filled using \texttt{maskfill}. 
It is important to note that these pixels are masked in all images due to NIRISS detector artifacts, such as those from the pick-off mirrors, rather than being part of a commonly source-masked region.

To enhance the quality and usability of the empirical backgrounds, it is necessary to smooth the resultant output. 
However, typical convolution-based smoothing methods tend to blur compact features present in the image, such as the 0$^{\rm th}$- and 1$^{\rm st}$-order dispersed pick-off mirror.
To address this, we use the \citet{darbonFastNonlocalFiltering2008} non-local means denoising algorithm implemented in \texttt{scikit-image}, which effectively smooths the background while preserving these compact features.

\begin{figure*}[ht!]
    \centering
    \includegraphics[width=\textwidth]{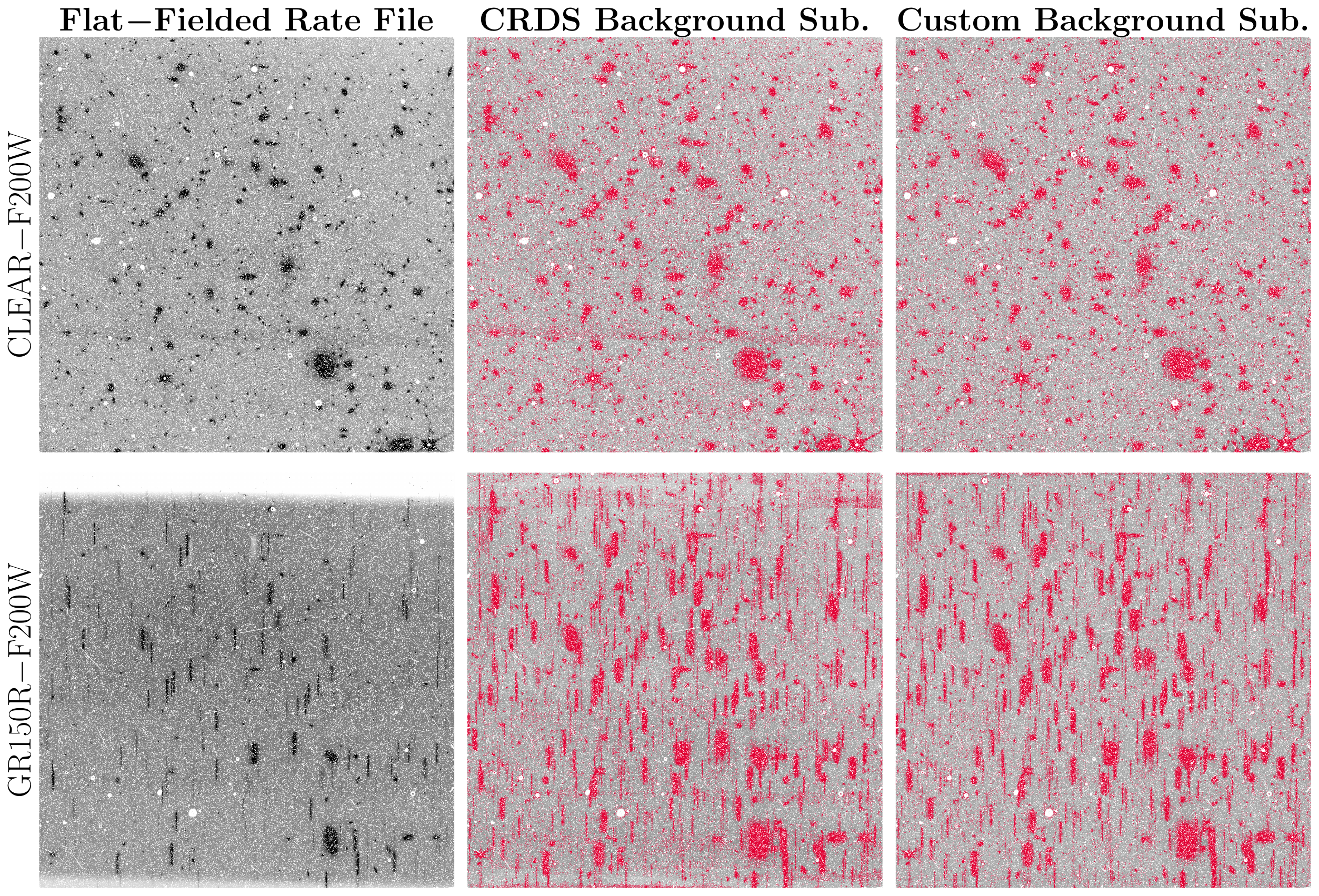}
    \caption{
        Comparison of source detection for a F200W/CLEAR (top) and  F200W/GR150R (bottom) image using the methodology described in Section \ref{subsec:mask_combine_smooth} with the CRDS reference file (middle) and the empirical background (right) \rev{with the input image  provided for comparison (left)}.
        Pixels with an active data quality flag are masked in white, while the source mask is shown in red.
        For imaging data, spurious sources are detected along the light saber, which is brighter than the background level.
        For WFSS data, the subtraction with the CRDS reference file results in spurious sources being detected, especially near the edges of the image, where the background subtraction residuals are most prominent.
        \label{fig:mask}
    }
\end{figure*}

\begin{figure*}[ht!]
    \centering
    \includegraphics[width=\textwidth]{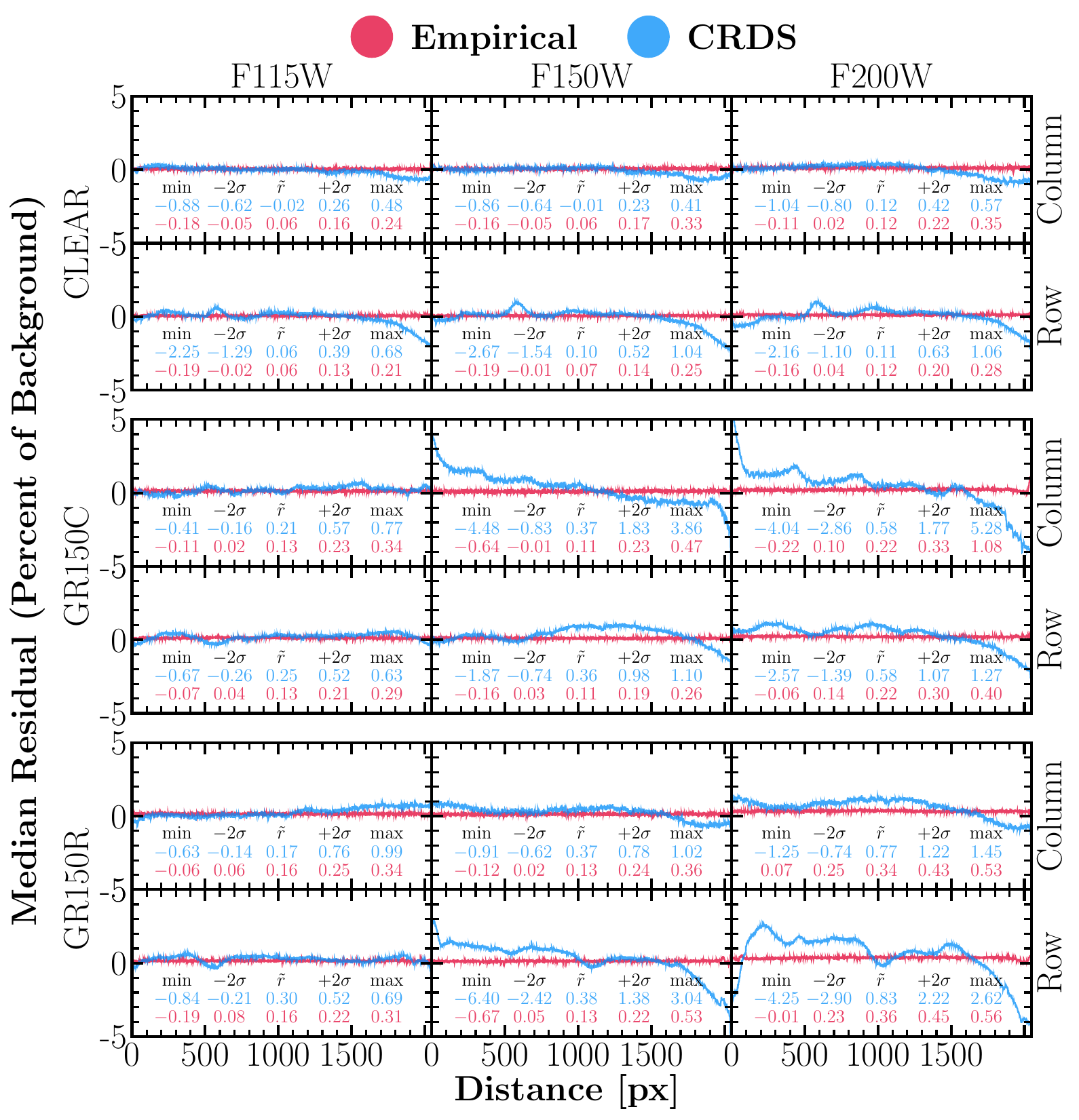}
    \caption{
        Median background subtraction residuals using the CRDS reference files (red) and the empirical backgrounds derived in this work (blue) across F115W (right column), F150W (middle column), and F200W (left column) filters and CLEAR (top row), GR150C (middle row), and GR150R (bottom row) grisms.
        Comparisons are made along detector columns (top inset) and rows (bottom inset) for each filter/grism combination.
        Residuals are presented as a percent of the total background level and as a function of detector position.
        We denote the 0, 5, 50, 95, and 100\% quantiles of the residuals in each comparison.
        \label{fig:sub}
    }
\end{figure*}

\subsection{Subtract \& Iterate} \label{subsec:subtract_iterate}

Once the smoothed background is generated, we subtract it from all images in the dataset.
The background is scaled to the original image using the \texttt{grizli} \texttt{visit\_grism\_sky} function which implements the \citet{brammerSourcedependentMasterSky2015} sky subtraction.
In short, an iterative approach is used to scale the smoothed background to the image where outliers to the fit, ideally from sources, are masked.
This deviates from the Stage 2 Spectroscopic pipeline which requires a detection catalog to mask sources before sky subtraction.
We prefer the approach implemented in \texttt{grizli} as it does not require a detection catalog and is independent of the grism dispersion model.
Finally, we subtract the scaled background from the original image to produce the final background-subtracted result.

Due to the dependence of the median-combined background on the masks used to exclude sources and artifacts, we pursue an iterative approach to refine the masks and improve the quality of the empirical backgrounds.
We therefore repeat the steps outline in Section \ref{subsec:mask_combine_smooth} using the final background-subtracted image to generate subsequent source masks.
In subsequent iterations we lower the source detection threshold to 1$\sigma$ above the background. 
These changes are made as the initial empirical background will have removed the majority of residual artifacts, enabling the detection of fainter sources in the image to improve the mask.

The iterative approach allows us to refine the masks and improve the quality of the empirical backgrounds.
A single additional iteration of background creation and source masking is sufficient to converge on a final background that is free from significant artifacts and accurately represents the background structure in the NIRISS imaging and WFSS data.
The flat-fielded and non-flat-fielded versions of our empirical backgrounds are publicly available at\dataset[doi:10.5281/zenodo.13838016]{https://doi.org/10.5281/zenodo.13838016} and are presented in Appendix \ref{app:backgrounds}.

\subsection{CRDS Subtraction} \label{subsec:crds}

To evaluate the effectiveness of our empirical backgrounds, we compare them to the subtractions performed with the CRDS reference files.
We use the CRDS reference files for WFSS data and a scalar value for imaging data as there are no existing backgrounds for NIRISS imaging data. 
These are subtracted from the original images following the same procedure as described in Section \ref{subsec:subtract_iterate} rather than the JWST pipeline approach to avoid the need for a detection catalog and ensure a consistent comparison.
We note that as CRDS provides non-flat-fielded backgrounds, we first apply the flat-field step to the CRDS reference files before subtraction.

\section{Results} \label{sec:results}

Following background subtraction using both the CRDS reference files and our empirical backgrounds, we compare the differences in the results to evaluate the effectiveness of our approach.
In this section, we present the results of our analysis, demonstrating the improvements in data quality achieved with our empirical backgrounds. 

\subsection{Background Levels} \label{subsec:zodi}

We first verify that our background subtraction is consistent with the expected zodiacal light contribution in NIRISS imaging and WFSS data.
We expect the overall background level, i.e.\ the median level and/or the scale value multiplied to the background, to be correlated with the absolute ecliptic latitude of the field.
Figure \ref{fig:zodi} shows the background level as a function of the absolute ecliptic latitude for all images in our dataset.
Background levels decrease with increasing absolute ecliptic latitude and with redder filter wavelengths, consistent with the results from \citet{rigbyHowDarkSky2023}.
We note several outliers in the data, which show enhanced background levels above expectations. 
These observations are all in dense fields, such as the Large Magellanic Cloud (LMC) and the Triangulum Galaxy (M33), where the background level is dominated by the target itself.

\subsection{Source Masking} \label{subsec:masking}

Accurate background subtraction is critical for source identification in NIRISS imaging and WFSS data. 
Typical source detection algorithms, such as \texttt{SEP} \citep{barbarySEPSourceExtractor2016}, rely on detecting pixels that are a certain threshold above the background level.
However, the presence of residual background structure can lead to spurious sources being detected.

In Figure \ref{fig:mask} we compare the source masking for a F200W/GR150R and a F200W/CLEAR image using the methodology described in Section \ref{subsec:mask_combine_smooth} the CRDS reference file and the empirical background \rev{with the input image  provided for comparison (left).}
For both imaging and WFSS data, standard subtraction using the CRDS reference files results in spurious sources being detected, especially along the light saber in imaging data and near the edges of the image in WFSS data where the background subtraction residuals are most prominent.
\rev{This occurs even with a low-frequency 2D background subtracted from the image as described in Section \ref{subsec:mask_combine_smooth}.}
Conversely, the empirical background is able to accurately mask sources without detecting spurious sources along the edges of the image or the light saber.

The differences in source masking between the CRDS reference files and our empirical highlight the importance of accurate background subtraction in NIRISS imaging and WFSS data.
\rev{We note that typically source detection is performed on a mosaic created from dithered exposures. 
While the exact impact depends on the overall background level, dither pattern, and mosaicing procedure, the combination of multiple exposures typically increases the number of areas in the resultant image in which spurious sources are detected, while decreasing the severity of the deviation of the image in these regions as the result is averaged from multiple inputs. 
However, $\sim$20\% of pure parallel visits have only one exposure in at least one direct imaging filter and are directly comparable to the example outlined in this section.}

\subsection{Background Subtraction Residuals} \label{subsec:resid}

To evaluate the effectiveness of our empirical backgrounds, we compare the residuals from background subtraction using the flat-fielded CRDS reference files and our empirical backgrounds.
We compute the median residuals using an iterative sigma-clipping algorithm with a clipping limit of 1$\sigma$ along detector columns and rows for each filter/grism combination across all images after excluding pixels in the respective source mask.

If the background subtraction is accurate, the residuals should be flat across the entire image, i.e.\ no trends should be observed in either spatial dimension.
In Figure \ref{fig:sub} we present the median residuals as a percent of the total background level and detector position.
Subtraction using CRDS reference files fail produce constant residuals across the entire image, with the residuals up to 6\% of the total background level present in the resultant image.
The effect is especially pronounced at the longest wavelengths, the edges of the image, and, for WFSS data, along the dispersion axis.
In addition, the residuals for imaging data show the influence of the light saber effect, with residuals at the $\sim$1\% level along the length of the detector.

Conversely, the empirical backgrounds produce residuals that are flat across the entire image.
We note that these residuals are not consistent with zero that one might expect from a perfect background subtraction, but rather are consistent with $\sim$0.06--0.3\% of the total background level.
We attribute this to that the source masks used do not capture faint sources or the outskirts of extended sources, leading to slight enhancement of the median residual.

Finally, residuals for both grisms and the F200W filter still show a slight artefact along the dispersion axis at the edge of the detector at the $\sim$0.25\% level which may signal additional image structure not captured solely by an empirical background.
We discuss the implications of this in Section \ref{sec:summary}.

\section{Summary \& Discussion} \label{sec:summary}

In this work we have constructed empirical backgrounds in the F115W, F150W, and  F200W filters for JWST NIRISS imaging and WFSS data using all public data as of \rev{2024-09-24} and make them available to the broader community at\dataset[doi:10.5281/zenodo.13838016]{https://doi.org/10.5281/zenodo.13838016}.
These backgrounds solve existing issues with existing CRDS reference images, which introduce spatially-dependent artifacts at the 1--5\% level in background-subtracted outputs.
In addition, for fields where the light saber is dominated by scattered zodiacal light, empirical background subtraction appears to be effective in mitigating the light saber effect, though this may require additional processing in the presence of bright stars in the susceptibility region.
Empirical background subtraction improves source identification in NIRISS imaging and WFSS data, a critical step in any subsequent astronomical analysis.
Our empirical backgrounds are publicly available and can be used to improve the quality of JWST NIRISS imaging and WFSS data reduction.
The code used to create these images is also publicly available \citep{hvidingTheSkyentistWFSSBackgroundsVersion2024a} and the methodology is broadly applicable to other NIRISS filters, modes, and potentially other JWST instruments.

However, we note that even with our empirical backgrounds there are remain artifacts after background subtraction for the GR150C and GR150R grisms in the F200W filter at the $\sim$0.25\% level.
These artifacts indicate that background subtraction alone may be insufficient to remove all residual structure in NIRISS imaging and WFSS data.
In this work we have assumed that all prior steps in the JWST pipeline are performed using calibration files that accurately represent the behavior of the detector, instrument, and telescope.
However, poorly calibrated flat-fields or other systematics can introduce artifacts similar to those seen in the residuals from background subtraction using the CRDS reference files.
Further work is needed to determine if the spatial structures seen in the NIRISS data are due to zodiacal backgrounds or are improperly removed by earlier stages of the JWST pipeline.


$ $\\ 


We would like to thank the anonymous reviewer for their constructive comments which improved the final manuscript.

This work makes use of color palettes created by Martin Krzywinski designed for colorblindness. The color palettes and more information can be found at \url{http://mkweb.bcgsc.ca/colorblind/}.

We acknowledge the support of the Data Science Group at the Max Planck Institute for Astronomy (MPIA) and especially M.\ Fouesneau for their assistance in developing methodology for this research paper.

This work is based in part on observations made with the NASA/ESA/CSA James Webb Space Telescope. The data were obtained from the Mikulski Archive for Space Telescopes at the Space Telescope Science Institute, which is operated by the Association of Universities for Research in Astronomy, Inc., under NASA contract NAS 5-03127 for JWST. These observations are associated with programs 1571, 3383, 4681.

The authors acknowledge the PASSAGE and OutThere teams for developing their observing program with a zero-exclusive-access period.

We acknowledge the support of the NIRISS team at the Space Telescope Science Institute for their feedback and assistance as a part of JWST Help Desk ticket INC0197167. 

This work has made use of data from the European Space Agency (ESA) mission Gaia (\url{https://www.cosmos.esa.int/gaia}), processed by the Gaia Data Processing and Analysis Consortium (DPAC, \url{https://www.cosmos.esa.int/web/gaia/dpac/consortium}).
Funding for the DPAC has been provided by national institutions, in particular the institutions participating in the Gaia Multilateral Agreement.


\facilities{JWST (NIRISS)}

\software{\texttt{Astropy} \citep{astropycollaborationAstropyCommunityPython2013}, \texttt{astroquery} \citep{ginsburgAstroqueryAstronomicalWebquerying2019}, \texttt{grizli} \citep{brammerGrizliGrismRedshift2019}, \LaTeX\ \citep{lamportLaTeXDocumentPreparation1994}, \texttt{Matplotlib} \citep{hunterMatplotlib2DGraphics2007}, \texttt{maskfill} \citep{vandokkumRobustSimpleMethod2024}, \texttt{NumPy} \citep{oliphantGuideNumPy2006,vanderwaltNumPyArrayStructure2011, harrisArrayProgrammingNumPy2020}, \texttt{photutils} \citep{bradleyAstropyPhotutils132024}, \texttt{scikit-image} \citep{scikit-image}, \texttt{SEP} \citep{barbarySEPSourceExtractor2016}}

\bibliographystyle{aasjournal}
\bibliography{bibliography}{}

\appendix
\restartappendixnumbering

\section{Empirical Backgrounds} \label{app:backgrounds}

\begin{figure*}[ht!]
    \centering
    \includegraphics[width=\textwidth]{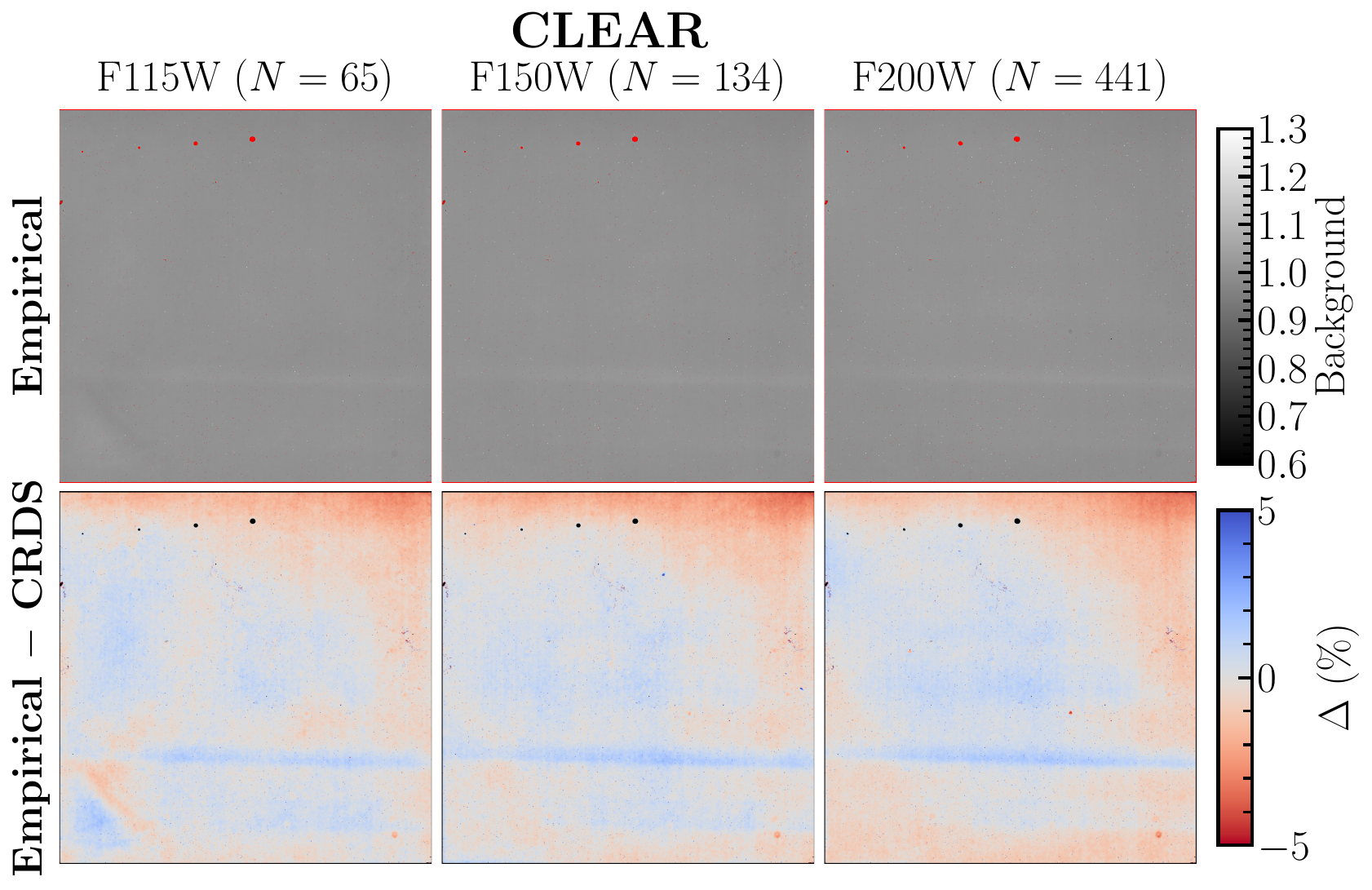}
    \caption{
        Empirical backgrounds (top row) and their difference from CRDS (bottom) for NIRISS imaging data in the F115W (left column), F150W (middle column), and F200W (right column) filters.
        Bad pixels are masked in red for the background images and in black for the difference images which include the bias regions and the occulting spots from the pick-off mirror.
        The number of contributing images is shown in the title of each column.
        As there are no existing backgrounds for NIRISS imaging data, we compare our empirical backgrounds to an array of ones as the `CRDS' reference.
        \label{fig:clear}
    }
\end{figure*}

\begin{figure*}[ht!]
    \centering
    \includegraphics[width=\textwidth]{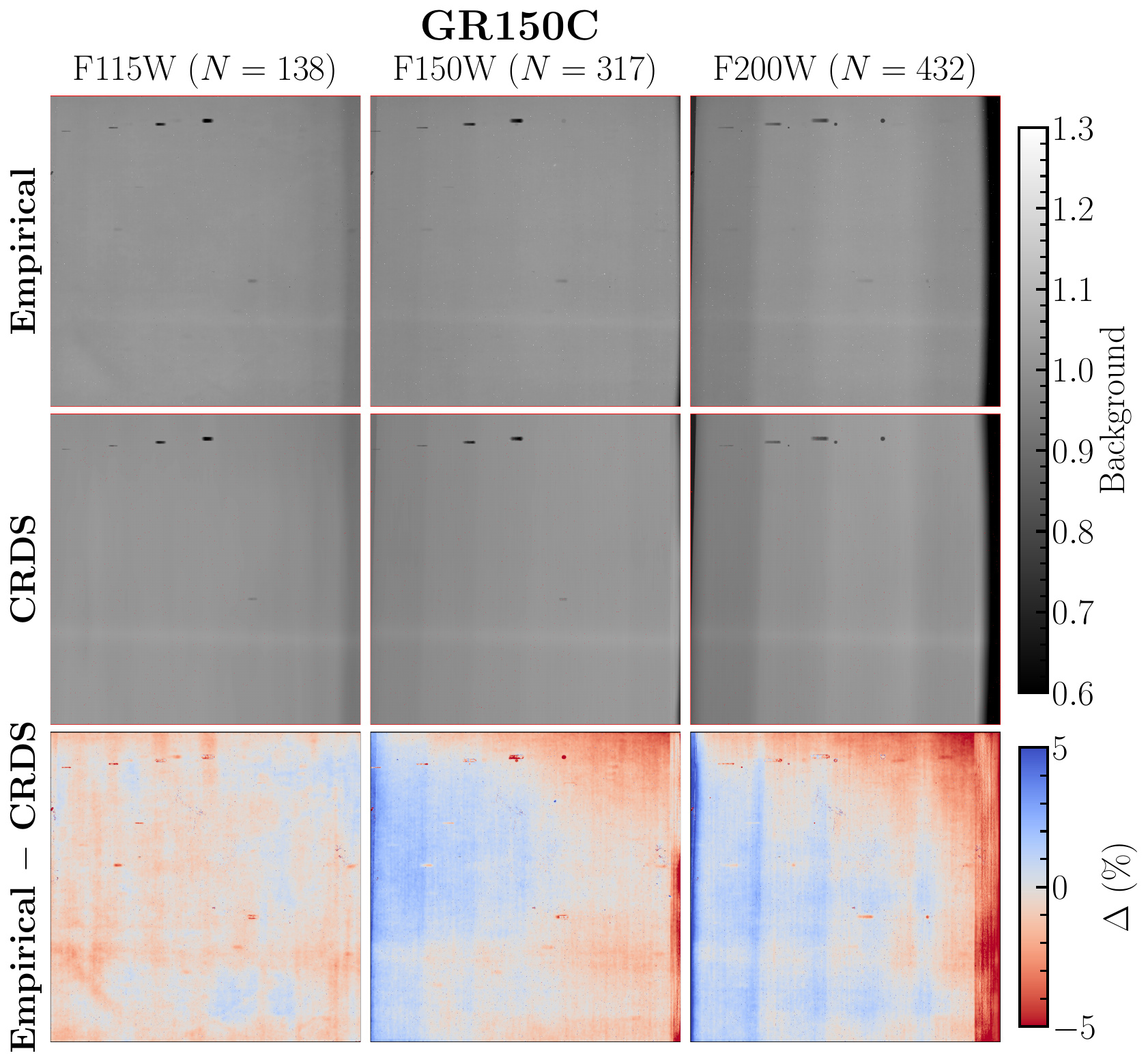}
    \caption{
        Empirical backgrounds (top row), flat-fielded CRDS reference files (middle row), and their difference (bottom row) for NIRISS WFSS data in the F115W (left column), F150W (middle column), and F200W (right column) filters with the GR150C grism.
        Bad pixels are masked in red for the background images and in black for the difference images which include the bias regions.
        The number of contributing images is shown in the title of each column.
        \label{fig:gr150c}
    }
\end{figure*}

\begin{figure*}[ht!]
    \centering
    \includegraphics[width=\textwidth]{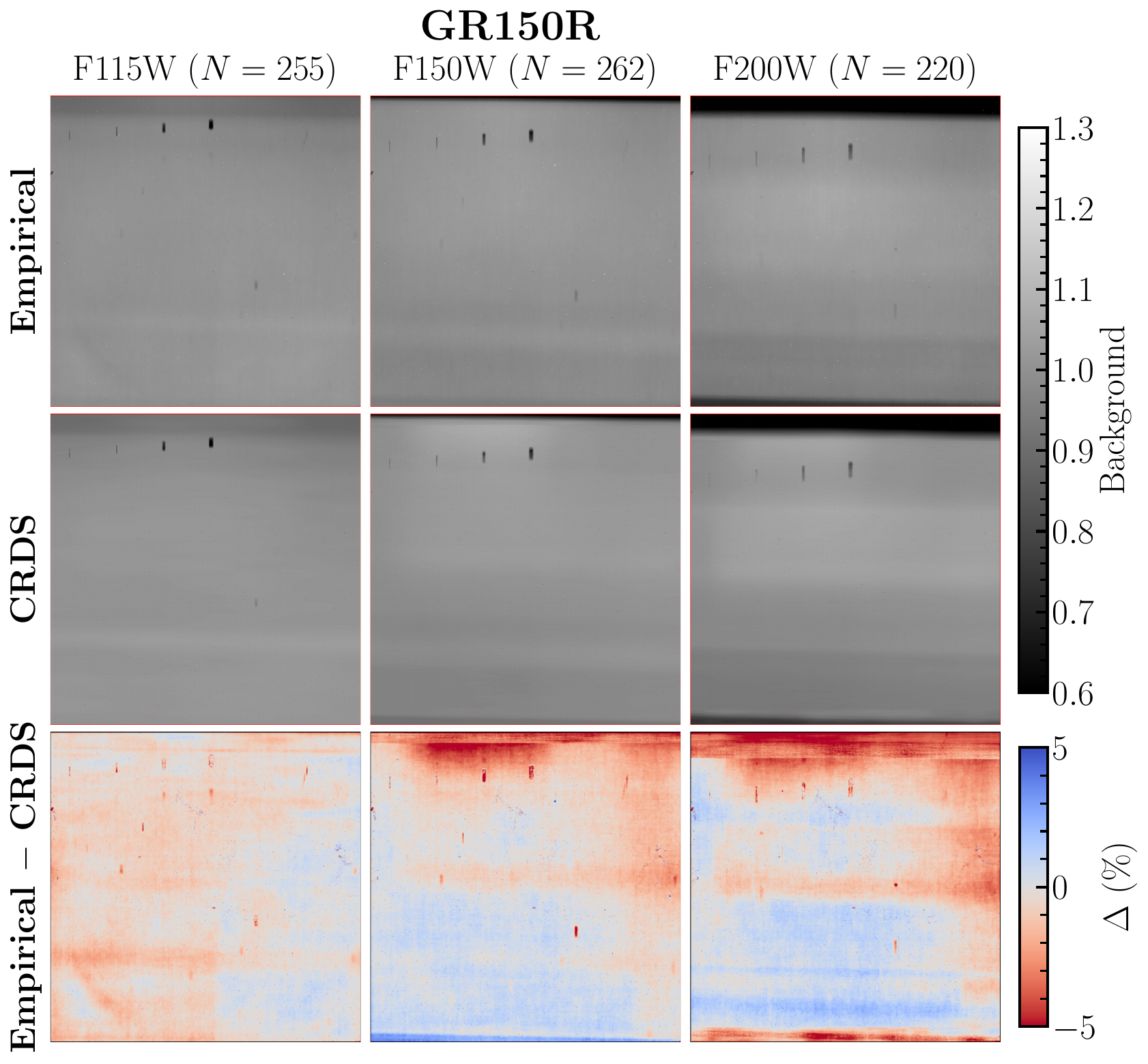}
    \caption{
        Empirical backgrounds (top row), flat-fielded CRDS reference files (middle row), and their difference (bottom row) for NIRISS WFSS data in the F115W (left column), F150W (middle column), and F200W (right column) filters with the GR150R grism.
        Bad pixels are masked in red for the background images and in black for the difference images which include the bias regions.
        The number of contributing images is shown in the title of each column.
        \label{fig:gr150r}
    }
\end{figure*}

In this appendix we present the flat-fielded versions of our empirical backgrounds for NIRISS imaging and WFSS data compared to the CRDS reference files available as of 2024-09-24 where available.
As CRDS WFSS backgrounds are not flat-fielded, we first apply the flat-field to the CRDS reference files before comparison.
NIRISS imaging backgrounds are presented in Figure \ref{fig:clear}, while WFSS backgrounds for GR150C and GR150R are presented in Figures \ref{fig:gr150c} and \ref{fig:gr150r} respectively.

We note that both imaging and WFSS backgrounds show 
In general, there is good agreement between the empirical backgrounds and the CRDS reference files, including the location and strength of the dispersed pick-off mirror occulting spots.
However, as discussed in Section \ref{sec:results}, there are significant differences at the $\sim$5\% level that are not distributed randomly: they are spatially dependent and can impact subsequent astronomical analysis.

Notably for imaging data across all filters the light saber appears at the $\sim$1\% level in the empirical backgrounds which is not accounted for by the JWST pipeline and CRDS reference files.
In general, differences are more pronounced in WFSS data for the redder filters where the CRDS reference files differ at the $\sim$5\% level.
These differences are most pronounced along the dispersion axis and at the edges of the detector.
The empirical backgrounds provided in this work can therefore be used to improve the quality of JWST NIRISS imaging and WFSS data reduction.

\end{document}